\documentclass[pdflatex,sn-apa]{sn-jnl}

\usepackage{hyperref}
\usepackage{url}
\usepackage{graphicx}%
\usepackage{multirow}%
\usepackage{amsmath,amssymb,amsfonts}%

\theoremstyle{thmstyleone}%
\theoremstyle{thmstyletwo}%

\theoremstyle{thmstylethree}%

\begin{document}

\title[Generative AI: Policy Impacts]{Adapting University Policies for Generative AI: Opportunities, Challenges, and Policy Solutions in Higher Education}

\author{\fnm{Russell} \sur{Beale}}\email{r.beale@bham.ac.uk}

\affil{\orgdiv{School of Computer Science}, \orgname{University of Birmingham}, \orgaddress{\street{Edgbaston}, \city{Birmingham}, \postcode{B15 2TT}, \country{UK}}}

\abstract{The rapid proliferation of generative artificial intelligence (AI) tools—especially large language models (LLMs) such as ChatGPT—has ushered in a transformative era in higher education. Universities in developed regions are increasingly integrating these technologies into research, teaching, and assessment. On one hand, LLMs can enhance productivity by streamlining literature reviews, facilitating idea generation, assisting with coding and data analysis, and even supporting grant proposal drafting. On the other hand, their use raises significant concerns regarding academic integrity, ethical boundaries, and equitable access. Recent empirical studies indicate that nearly 47\% of students use LLMs in their coursework—with 39\% using them for exam questions and 7\% for entire assignments—while detection tools currently achieve around 88\% accuracy, leaving a 12\% error margin. This article critically examines the opportunities offered by generative AI, explores the multifaceted challenges it poses, and outlines robust policy solutions. Emphasis is placed on redesigning assessments to be AI-resilient, enhancing staff and student training, implementing multi-layered enforcement mechanisms, and defining acceptable use . By synthesizing data from recent research and case studies, the article argues that proactive policy adaptation is imperative to harness AI’s potential while safeguarding the core values of academic integrity and equity.}

\keywords{Generative AI, Large Language Models, Higher Education, Academic Policy, Academic Integrity, Teaching Innovation, Research Productivity}



\maketitle

\section{Introduction}\label{sec1}

The advent of generative AI, most notably exemplified by large language models (LLMs) like ChatGPT, marks a pivotal moment in the evolution of higher education. Since its public introduction in late 2022, ChatGPT has rapidly gained traction among researchers, educators, and students alike. Its ability to review literature, produce human-like text, generate code, and  synthesise complex ideas has led to widespread experimentation across academic disciplines.

Universities in Australia, New Zealand, the UK, US, and other developed nations are now grappling with the implications of this technology. On the research front, LLMs are used to accelerate literature reviews, refine research questions, and even draft grant proposals. In the classroom, instructors employ AI-powered virtual teaching assistants, adaptive learning modules, and personalised feedback systems. However, the widespread use of these tools also brings significant challenges, particularly in assessment practices and the maintenance of academic integrity.

This article provides an examination of the opportunities and challenges generated by LLMs in academic contexts. It is not a comprehensive review of all LLM approaches in higher education: rather, it presents the key issues distilled from the field,  highlights key empirical data regarding usage rates, disciplinary differences, and detection tool performance, and then focuses on the critical policy issues that institutions must address. The central argument is that while generative AI offers transformative potential for teaching and research, robust and adaptive policies are essential to mitigate risks and uphold academic standards.  The philosophical question posed is, since the tools now produce simulacrums of knowledge, how can we teach understanding and enquiry - a version of the age-old question, what are Universities for?

The structure of the article is as follows. Section \ref{opportunities} outlines the opportunities that generative AI presents for research and teaching. Section \ref{challenges} examines the challenges and risks associated with AI integration—focusing on misuse in assessments, detection limitations, and ethical concerns.  Section \ref{case studies} discusses the imperative for adaptive policies and reviews institutional responses and case studies. Section \ref{discussion} synthesises the discussion and offers concrete policy recommendations. Finally, Section \ref{conclusion} provides concluding reflections and future directions.

The research articles and materials for this were retrieved via relevant searches on Google Scholar, Google, following a lightweight prisma methodology \citep{page_prisma_2021} and via prompt engineered conversations with ChatGPT-o3-mini-hi, and 4.0. Generative AI surveyed parts of the literature and identified areas for exploration; careful prompt engineering focussed the work to provide the basic themes, which have been shaped, written and owned by the author.  The key contribution proposing to make the most significant impact by shaping the curriculum and assessment is the author's.  

\section{Opportunities Presented by Generative AI in Academia}\label{opportunities}

Generative AI holds immense promise for higher education. Its applications span the entire academic spectrum in all subjects, from research to teaching, and have the potential to revolutionise traditional practices. In this section, we detail the principal opportunities that LLMs offer, supported by empirical evidence and case studies.

\subsection{Enhancing Research Productivity}

\subsubsection{Accelerated Literature Reviews and Knowledge Synthesis}
One of the most time-consuming tasks in research is the literature review. Traditionally, scholars sift through vast databases, identify key articles, and manually synthesise findings. LLMs, however, can rapidly process large volumes of text and generate concise summaries. For instance, \citet{Tang2024} evaluated the performance of LLMs in generating literature reviews. Their study revealed that while LLMs can produce coherent and structured summaries, they are prone to generating “hallucinated” references—that is, plausible but incorrect citations. This finding underscores the need for human oversight. Nevertheless, the potential efficiency gains are substantial. By using LLMs as an initial drafting tool, researchers can significantly reduce the time spent on background research, allowing them to focus on analysis and interpretation.  Newly released 'deep knowledge' options allow for further, deeper explorations of topics with the LLM providing a structured response, and some models show their reasoning around the search strategies. allowing for validation of the approach.

\subsubsection{Facilitating Idea Generation and Refinement of Research Questions}
Beyond summarizing existing literature, LLMs serve as effective brainstorming partners. Researchers can interact with these models to explore novel research directions, refine existing questions, and identify gaps in the literature. \citet{Korinek2023} discusses how generative AI can stimulate the ideation process in economic research by automating routine cognitive tasks. The ability to rapidly generate alternative hypotheses or suggest interdisciplinary connections is particularly valuable in fields where innovation is key. Although the output of LLMs must always be critically assessed, the capacity for generating creative insights is an undeniable advantage.

\subsubsection{Supporting Grant Proposal and Manuscript Drafting}
Grant proposals and research manuscripts require clear, persuasive writing—a task that many researchers find challenging, particularly those for whom English is a second language. \citet{Seckel2024} provide guidelines for leveraging LLMs in drafting grant proposals. They caution that while AI can improve language clarity and suggest structure, the final narrative must reflect the researcher’s own intellectual contributions. In practice, LLMs have been used to generate first drafts of research papers and grant applications, which are then refined and verified by human experts. This collaborative process not only expedites writing but also reduces the likelihood of errors due to language barriers.

\subsubsection{Assisting in Data Analysis and Coding}
In STEM disciplines, the ability of LLMs to generate code and assist with data analysis is particularly impactful. Many researchers face bottlenecks due to the time required to write and debug code. LLMs, integrated into platforms like GitHub Copilot, can generate code snippets and offer debugging support, thereby accelerating the research process. Korinek \citep{Korinek2023} notes that such tools democratise access to advanced computational methods, enabling researchers with limited programming expertise to conduct complex analyses. This assistance is not without risks—errors in code or misinterpretations of statistical outputs can occur—but the potential for increased productivity is significant.

\subsection{Transforming Teaching and Learning}

\subsubsection{Virtual Teaching Assistants and Automated Support}
The teaching landscape is also being reshaped by generative AI. In large-enrollment courses, the provision of timely, individualised support is a perennial challenge. AI-powered virtual teaching assistants (VTAs) are emerging as a solution. Systems such as “JeepyTA” \citep{shah_students_2024} have been deployed in computer science courses to answer frequently asked questions around the clock. LLM-powered chatbots can provide answers to common questions ,and direct students to resources and other forms of support, freeing up human assistance for more complex tasks \citep{labadze_role_2023}. The use of VTAs has been associated with reduced response times and increased student satisfaction. Importantly, these tools also support accessibility by providing immediate assistance outside traditional office hours.

\subsubsection{Adaptive Learning and Personalised Feedback}
Generative AI has the potential to create highly personalised learning experiences \citep{razafinirina_pedagogical_2024}. By analyzing student performance data, LLMs can generate adaptive feedback that is tailored to individual needs. In a controlled study, \citet{Kinder2024} demonstrated that pre-service teachers who received adaptive, AI-generated feedback showed significant improvements in the quality of their written justifications. Although the study noted that decision accuracy remained unchanged, the increased engagement and depth of reflection are promising indicators of the tool’s pedagogical value. Personalised feedback helps students identify their weaknesses and build on their strengths—a critical factor in effective learning.   \citet{bloom_2_1984} noted  that the average student tutored one-to-one could perform two standard deviations better than students educated in a conventional manner. i.e. personal tutoring provides an average 98\% above the level of their colleagues, and 90\% of the students tutored this way achieved results that only 20\% of conventionally taught students could achieve.  One-to-one tuition can make everyone a prodigy: we cannot ignore its potential to help all students.

\subsubsection{Content Creation and Curriculum Enrichment}
Instructors can also leverage LLMs to enhance their teaching materials. The ability of AI to generate diverse instructional content—including lecture outlines, case studies, and quiz questions—enables educators to quickly adapt materials for different learning levels and contexts. For example, an instructor can request an introductory explanation of a complex concept for first-year students and a more advanced treatment for upper-level courses. This dynamic content generation supports differentiated instruction and allows for rapid curriculum updates in response to emerging trends and discoveries.

\subsubsection{Enhancing Student Engagement and Inclusive Learning}
Generative AI tools have the potential to increase student engagement and promote inclusive learning environments. Studies have shown that when AI-generated content is used as a basis for class discussions, students are more likely to engage critically with the material. Furthermore, AI tools can be particularly beneficial for students who face language barriers or learning disabilities. By providing simplified explanations, translations, or alternative perspectives, LLMs can help ensure that all students have access to high-quality educational resources. The \citet{HEPI2025} survey indicates that 67\% of UK students view AI as essential, suggesting that these tools are rapidly becoming a core component of the learning process.

\section{Challenges and Risks of Integrating Generative AI in Academia}\label{challenges}
Despite the considerable opportunities presented by generative AI, its integration into academic environments is not without significant challenges. In this section, we explore the risks associated with its use—particularly in assessments and academic integrity—and discuss the implications for diverse academic disciplines.

\subsection{Misuse in Student Assessments}
\subsubsection{Prevalence of AI-Assisted Work}
Recent empirical studies reveal a concerning trend regarding the misuse of generative AI in academic assessments\citep{Kim2025}. \citet{Paustian2024} report that approximately 46.9\% of students use LLMs in their coursework. More disturbingly, 39\% of surveyed students admitted to using AI tools to answer exam or quiz questions, while 7\% confessed to having employed AI to write entire assignments. These figures suggest that the convenience and efficiency of LLMs are enticing students to bypass the genuine learning process. The implication is that academic assessments may no longer accurately reflect a student’s own understanding or abilities if AI-generated work is not properly disclosed.

\subsubsection{Limitations of AI Detection Technologies}
In an effort to counteract academic misconduct, a range of AI-detection tools have been developed. However, these technologies are not yet foolproof. Studies indicate that the current generation of AI detectors achieves an accuracy rate of approximately 88\%, meaning that about 12\% of AI-generated content may go undetected \citep{Paustian2024}. This error margin is particularly concerning in high-stakes assessments where even a small number of false negatives or false positives can have significant consequences. These detection tools often work by detecting stylistic patterns and limitations of variance of languages use in submitted work, which whilst characteristic of LLM generated text is also more likely in non-native English speakers, risking adverse discrimination n the detections. Reliance solely on automated detection systems is insufficient; a more comprehensive approach is required that combines technological and human oversight.

\subsection{Disciplinary and Demographic Variations in AI Adoption}
\subsubsection{Variability Across Academic Disciplines}
Generative AI adoption is not uniform across all fields of study. Research \citep{Kim2025} highlights that students in STEM and Health-related disciplines are more inclined to use AI tools compared to those in the humanities and arts. In STEM fields, assignments often involve problem-solving and coding tasks, which align naturally with the capabilities of LLMs. Conversely, in humanities disciplines—where critical analysis and original thought are highly valued—there is greater caution regarding AI use. Concerns over originality and the potential dilution of interpretative depth lead to lower adoption rates. This disciplinary divide necessitates that policies be tailored to address the specific challenges and benefits experienced by different academic fields.

\subsubsection{Socioeconomic and Gender Disparities}
In addition to disciplinary differences, socioeconomic and gender gaps in AI usage have emerged. The HEPI (2025) survey \citep{HEPI2025} found that male students and those from higher socioeconomic backgrounds are significantly more likely to use generative AI tools than female students or those from lower socioeconomic strata. Such disparities not only reflect differences in technological familiarity and access but also risk exacerbating existing educational inequalities. Policies must therefore consider mechanisms to ensure equitable access to AI tools and training, thereby preventing a digital divide from undermining academic fairness.

\subsection{Ethical and Pedagogical Concerns}
\subsubsection{Erosion of Academic Integrity}
The potential for AI to generate high-quality, human-like text raises critical questions about academic integrity. When students use generative AI to produce assignments without proper disclosure, the work submitted bears less of a direct relationship to their own intellectual efforts. The link between submitted work and student comprehension is therefore broken, and this undermines the assessment process and devalues the credentials awarded by academic institutions. In environments where original thought and critical engagement are prised, unchecked AI use can lead to a dilution of academic standards. \citet{Cotton2024} argue that universities must address these challenges head-on by revising honour codes and explicitly defining what constitutes unacceptable use of AI.

\subsubsection{The Black-Box Problem and Lack of Transparency}
Another significant challenge is the opaque nature of LLMs. Despite their impressive outputs, the inner workings of these models remain largely inscrutable. This “black-box” phenomenon complicates efforts to verify the accuracy and originality of AI-generated content. Educators are thus forced to contend not only with potential errors and fabricated references \citep{huang_survey_2025} but also with the inherent uncertainty of relying on technology whose decision-making processes are not fully understood. Such opacity calls for policies that mandate rigorous verification and emphasise human oversight, ensuring that AI outputs are always cross-checked against reliable sources.  More recent models are getting better at reducing hallucinations and making their reasoning processes visible, and multi-shot interactions \citep{dhuliawala_chain--verification_2023} and careful prompting \citep{phoenix2024prompt, marvin_prompt_2024}  are also improving reliability and reproducibility but this still currently requires expert knowledge.

\subsubsection{Bias, Fairness, and the Risk of Misinformation}
Generative AI systems are trained on vast datasets that may contain biases. As a result, the outputs of these models can inadvertently reflect or amplify existing prejudices (for example, current diffusion models for image creation cannot create left-handed people). In academic settings, biased or misleading information can have profound implications—both for research integrity and for student learning \citep{mehrabi_survey_2021}. Researchers and educators must be vigilant in detecting such biases, and policies should include provisions for regular audits of AI-generated content. Ensuring fairness and accuracy is essential not only for maintaining academic standards but also for fostering an inclusive learning environment.

\subsubsection{Trust, Accountability, and the Role of Human Oversight}
The rapid adoption of generative AI has outpaced the development of comprehensive accountability frameworks. Trust in AI tools is inherently tied to the ability of institutions to hold them accountable for errors or misuse. The current reliance on detection tools—with their known limitations—highlights the need for an integrated approach that blends technological solutions with manual review processes. Policies must clearly articulate the roles and responsibilities of both AI systems and human agents in maintaining academic integrity.

\section{Does LLM use actually reduce understanding?}
There is a prevailing perspective that unfettered use of LLMs reduces understanding and hence is A Bad Thing, though the evidence for this is scarce.  Certainly, AI can produce rapid solutions to problems previously thought to be challenging: the author created a solution to a Computer Science Masters project that would easily have passed, and instead of the 5 months the project was supposed to take, the solution was created, with copious assistance from an AI coding assistant, in 8 minutes.  But this is more a case of the project being inappropriate nowadays than anything else. 

Research and educator reports suggest that if students over-rely on AI-generated results (using it as a shortcut rather than a learning aid), they \textit{might} not fully develop critical  skills, such as concept understanding, structured thinking, and problem-solving. This can lead to a more superficial grasp of underlying concepts. Conversely, when used appropriately—as a supplement to traditional instruction—LLMs can enhance learning by providing instant feedback, clarifying  concepts, and even demonstrating multiple approaches to a problem. For instance, using AI as an interactive tutor or as part of a guided  exercise can help students compare their own solutions with AI-generated ones, deepening their understanding.  In summary, the risk of decreased understanding is real if LLMs are used just to fill assessment needs. However, with careful pedagogical design that emphasises critical evaluation and hands-on practice, these tools have the potential to enhance—not diminish—students' coding comprehension.

\section{The Imperative for Adaptive Policy in Higher Education}\label{adaptive policy}
The challenges outlined above underscore the critical need for adaptive policies that both leverage the benefits of generative AI and mitigate its risks \citep{Kim2025}. Universities must develop comprehensive frameworks that address ethical, pedagogical, and technological dimensions. In this section, we outline the key components of such adaptive policies and review institutional responses that offer promising models.

\subsection{Defining Acceptable Use of Generative AI}
A cornerstone of effective policy is the clear delineation of acceptable versus unacceptable uses of generative AI in academic work. Institutions should adopt guidelines that:

\textbf{Clarify Permitted Applications:} Define specific contexts in which AI assistance is allowed (e.g., for initial drafting, language editing, brainstorming) while prohibiting practices that outsource entire assignments or exam responses.

\textbf{Mandate Transparency and Disclosure:} Require that any AI-generated content be explicitly disclosed in research publications, grant proposals, and coursework. This mirrors traditional citation practices and reinforces accountability.

\textbf{Provide Illustrative Examples:} Use case studies and concrete examples to clarify the boundaries of ethical AI use. By illustrating acceptable and unacceptable practices, institutions can reduce ambiguity and guide student behaviour.

These are good principles and set a values-based agenda for students to understand and follow.  By themselves, however, they are inadequate, as they do not provide specific, actionable constructs for educators or student alike, and are open to abuse.

\subsection{Redesigning Assessments to Foster Originality}
Traditional assessment formats—such as take-home essays—are particularly vulnerable to AI misuse. Adaptive policies should promote assessment designs that emphasise the process of learning as much as the final product. Recommended approaches include:
\\
\textbf{Real-Time Assessments:} Utilise in-class examinations, oral defenses, and timed assessments that require spontaneous, original responses.
\\
\textbf{Process Documentation:} Require the submission of drafts, work logs, and reflective narratives that document the student’s thought process and revision history. This approach not only deters AI misuse but also rewards genuine intellectual effort.
\\
\textbf{Collaborative and Peer-Reviewed Projects:} Design assignments that incorporate group work and peer review, where individual contributions can be more easily validated.

\textbf{Explanation-based assessment}: in which the content can be created using any tools, including AI ones. but the student is required to explain the content, discuss why that approach was taken and what alternatives were explored. and so on.  for sure, all these issues could ne explored beforehand with AI assistance, but if the student knew all of this information, it's likely that they have sufficient understanding.
\\
\textbf{Scenario-Based and Application-Oriented Tasks:} Develop assessment items that require the application of knowledge to novel scenarios—tasks that are less amenable to generic AI responses.

\subsection{Enhancing Training and AI Literacy}
For policies to be effective, all academic stakeholders must be equipped with the skills and knowledge necessary to navigate an AI-enhanced educational landscape. This includes:
\\
\textbf{Staff Development Programs:} Organise regular workshops and training sessions for instructors on the ethical use of AI, techniques for detecting AI-generated content, and methods for designing AI-resilient assessments. Such programs can also cover best practices for integrating AI tools into teaching.\\
\textbf{Student Orientation and Ongoing Support:} Integrate modules on AI literacy into first-year orientations and offer refresher courses throughout the academic year. Topics should include responsible use, proper attribution, and the limitations of AI-generated information.\\
\textbf{Interdisciplinary Forums and Communities of Practice:} Establish forums where educators across disciplines can share experiences, discuss challenges, and collaboratively refine institutional policies. These communities can serve as incubators for innovative teaching practices and adaptive policy frameworks.

However, in practice, most initiatives currently focus on introductory awareness—covering basic concepts, ethical considerations, and the limitations of generative AI—rather than offering comprehensive, advanced technical training.  The \citet{HEPI2025} survvey of UK universities found that while 67\% of students consider AI essential to their academic work, only a limited portion have participated in in-depth AI training. This suggests that many courses remain at a basic level of awareness. Although both staff and students recognise the importance of generative AI in modern education, only about 30–40\% of respondents have received formal training on these tools \citep{Kim2025}. This highlights a gap between the recognised need for AI literacy and the current provision of advanced, hands-on courses.

\subsection{Multi-Layered Enforcement and Accountability}
Given the limitations of current AI detection technologies, having assessments that are amenable to generative AI answering are the fundamental issue.  If these cannot be altered, or are offered as part of a suite of assessment approaches, enforcement of academic policies must rely on a combination of automated and manual oversight:
\\
\textbf{Integration of Detection Tools with Human Review:} Use AI-detection software as an initial screening mechanism, but ensure that flagged cases are reviewed by trained academic integrity officers.\\
\textbf{Regular Policy Audits and Updates:} Establish mechanisms for periodic review of AI-related policies, ensuring that they evolve in response to technological advancements and emerging usage patterns.\\
\textbf{honour Codes and Self-Reporting Initiatives:} Encourage a culture of academic integrity by promoting honour codes that explicitly address the use of generative AI. Systems that allow for self-reporting and remediation can foster a more ethical academic environment.\\
\textbf{Transparent Reporting and Feedback Loops:} Implement processes that allow for feedback on the effectiveness of enforcement measures, and ensure that policy revisions are informed by empirical data and stakeholder input.
\textbf{Student involvement and feedback:} Processes that encourage student reflection and analysis on the successes or otherwise of different teaching and assessment methods using LLMs should be encouraged, allowing them to have a strong voice in the development of good practice.

\section{Institutional Responses and International Case Studies}\label{case studies}
The response to the challenges posed by generative AI has varied across institutions and regions. This section highlights several case studies and institutional initiatives that provide models for effective policy adaptation.

Policies are enacted at individual levels across Australian institutions, with TEQSA providing a collection of useful resources \citep{noauthor_artificial_nodate}, and the general approach is in line with that of other countries.  However, whilst there are some general principles - don't use it to submit assessed work, do use it to help learn, these are not specifically helpful.  Monash provides a good example of encouraging higher level thinking amongst academics to drive a move towards GenAI-appropriate assessments, highlighting some of the considerations discussed above \citep{monash_university_ai_nodate}.  In the UK, the Russell Group — a coalition of 24 research-intensive UK universities — has been at the forefront of formulating policies to guide AI use in academia.  Their policy \cite{noauthor_russell_2023} can be summarised as a commitment to ethical use across teaching, learning, and assessment; building AI literacy, support and training for staff and students; evolving teaching and assessment ensuring that access to AI tools is fair and does not compromise learning outcomes, upholding academic integrity by defining when and where generative AI can be used; and sharing of best practice.  Individual universities have since developed detailed guidelines mandating disclosure of AI use and clarifying acceptable practices. These appear at first sight to be appropriate - a typical example would be:
\begin{quote}
\begin{itemize}
    \item you cannot use the output of Generative AI (i.e., the content it creates) in any assessment, unless explicitly authorised.
    \item While Generative AI can be a valuable tool for understanding concepts and exploring ideas, it should never replace your own critical thinking and analysis. Use it as a supplement to enhance your understanding, rather than as a substitute for your own work.
    \item These tools should not be used to write your assignment or theses and/or to clarify and develop your arguments. You should not use them to translate large amounts of text into English.

\end{itemize}
\end{quote}
These principles tend to be the extent of policy across almost all Universities worldwide at present.  But the reality is that this doesn't translate into actionable events when applied to students and courses: you can use AI to help, but only so far.  But exactly how far, and how it can be evaluated if that boundary has been crossed, is manifestly unclear.
\\
More specific, achievable changes have been taken:
\textbf{Revised Assessment Formats:} Several member institutions have piloted alternative assessment methods. These include in-class exams, oral defences, and project-based evaluations designed to reduce the opportunity for AI misuse.\\
\textbf{Staff and Student Training:} Many Russell Group universities have instituted  training programs to build AI literacy among staff and students. Workshops, online modules, and interdisciplinary forums have been deployed to ensure that all stakeholders understand both the potential and the limitations of generative AI.

U.S. universities have also taken proactive steps in response to the AI revolution. Notable initiatives include:
\\
\textbf{Revised Academic Integrity Policies:} Institutions such as Stanford, MIT, and the University of California system have updated their academic integrity guidelines to include explicit clauses on the use of generative AI. These policies emphasise transparency, proper attribution, and the consequences of misuse.  Again. however, the policing of this in practice remains unclear\\
\textbf{Innovative Teaching Practices:} U.S. institutions have integrated virtual teaching assistants and AI-powered tutoring systems into their courses. Early studies indicate that these tools can improve student engagement and reduce instructor workload, though challenges remain in verifying the authenticity of student work.\\
\textbf{Hybrid Detection and Enforcement:} Recognizing the limitations of automated AI detectors, many U.S. universities have adopted a hybrid approach that combines detection software with manual review by academic committees. This multi-layered strategy aims to balance efficiency with fairness in enforcement.

While much of the current literature focuses on the UK and US, other international perspectives also provide valuable insights:
\\
\textbf{European Initiatives:} Several European universities have begun collaborative projects aimed at standardizing AI policy across borders. These initiatives emphasise cross-national dialogue and the sharing of best practices to address the ethical and practical challenges of generative AI.\\
\textbf{Asian Case Studies}: Preliminary reports from parts of Asia suggest that while adoption rates may vary, the fundamental challenges—particularly around academic integrity—are universal. Institutions in these regions are also exploring adaptive assessments and enhanced training programs as part of their policy responses.

\section{Discussion: Balancing Innovation with Integrity}\label{discussion}
The dual nature of generative AI’s impact in higher education necessitates a balanced, nuanced approach. On one hand, the transformative potential of LLMs to enhance research productivity and teaching efficacy is clear. On the other, unchecked use poses serious risks to academic standards, fairness, and originality. The biggest issue is around its unauthorised use in assessment, which has the capability of undermining trust in the integrity of University qualifications.  This discussion section synthesises the opportunities and challenges outlined above and examines the broader implications for institutional culture and future policy development.

\subsection{Reconciling Efficiency Gains with Ethical Obligations}
The promise of generative AI lies in its ability to handle routine tasks—accelerating literature reviews, generating draft content, and even coding. However, these efficiency gains come at a cost if they lead to superficial learning or academic dishonesty. Institutions must therefore reconcile the push for productivity with the need to nurture critical thinking, originality, and ethical behaviour. Transparent disclosure and rigourous verification are key elements in ensuring that AI remains a tool for enhancement rather than a shortcut to academic success.

\subsection{The Role of Institutional Leadership and Culture}
Effective policy adaptation is not solely a technical or procedural challenge—it is also a matter of institutional culture and leadership. Universities that foster a culture of integrity and continuous learning are better positioned to integrate AI responsibly. Leadership must invest in training, provide clear communication about policy changes, and create forums for ongoing dialogue. Only through a collaborative approach can institutions ensure that AI tools are used to complement, rather than compromise, academic rigour.

\subsection{Future Directions: Research and Policy Innovation}
Looking forward, several research directions are crucial for informing policy innovation:
\\
\textbf{Longitudinal Impact Studies:} Continued research is needed to assess the long-term effects of generative AI on student learning outcomes and academic performance. Such studies should track changes over multiple academic cycles.\\
\textbf{Comparative Analyses Across Disciplines:} Given the disciplinary variations in AI adoption, comparative research can identify best practices that are adaptable across different fields.\\
\textbf{Technological Advances in Detection:} As AI models evolve, so too must the tools for detecting AI-generated content. Ongoing research into more accurate, context-sensitive detection algorithms is essential.\\
\textbf{Ethical Frameworks for AI Use:} Interdisciplinary work involving ethicists, educators, and technologists can help develop robust ethical frameworks that guide both policy and practice.

\section{Policy Recommendations}\label{policy}

Drawing on the empirical evidence and case studies discussed, the following policy recommendations are proposed to guide universities in integrating generative AI responsibly:

\begin{enumerate}

\item \textbf{Redesign Assessments to Emphasise Process and Originality}
   \begin{enumerate}
     \item \textit{Real-Time and In-Class Assessments:} Emphasise assessments that require immediate, unscripted responses. Oral exams, in-class writing tasks, and scenario-based assessments are less susceptible to AI misuse, regardless of how the materials relied on in that session were created.     
    \item \textit{Process Documentation:} Require students to submit drafts, annotated work logs, and reflective narratives that document their learning process and development of ideas.
    \item \textit{Collaborative Projects:} Design assignments that involve group work and peer evaluation, ensuring that individual contributions are evident and verifiable.
    \item \textit{Application-Based Tasks:} Focus on assessments that require students to apply concepts to novel scenarios, which are less likely to be answered by generic AI output.

\end{enumerate}

\item \textbf{Enhance Training and AI Literacy for Staff and Students}
    \begin{enumerate}
        \item \textit{Staff Development:} Organise regular workshops and training sessions on the ethical use of AI, techniques for detecting AI-generated work, and strategies for integrating AI into teaching.
        \item \textit{Student Orientation:} Include comprehensive modules on AI literacy in student orientations and first-year courses. These should cover ethical considerations, proper citation practices, and the limitations of generative AI
\item \textit{Interdisciplinary Forums:} Establish committees or forums that include representatives from different academic disciplines to share best practices and continuously update policies in light of technological advancements.
\end{enumerate}

\item \textbf{Implement Multi-Layered Enforcement and Detection Mechanisms}
    \begin{enumerate}
        \item \textit{Hybrid Detection Systems:} Utilise AI-detection tools as an initial filter, complemented by thorough manual reviews conducted by trained academic integrity officers.
        \item \textit{Regular Policy Audits: }Institutions should schedule periodic reviews of AI-related policies and detection methodologies to ensure they remain current with technological progress.
        \item \textit{Honour Codes and Self-Reporting:} Develop and promote honour codes that specifically address AI use, and create mechanisms that encourage self-reporting and remediation rather than immediate punitive measures.
        \item \textit{Transparent Feedback Loops: }Establish systems that allow for regular feedback from staff and students on the effectiveness of enforcement measures, ensuring that policies can be refined based on the other real-world experience.
        \end{enumerate}

\item \textbf{Develop Clear and Detailed AI Usage Guidelines}
\begin{enumerate}
        \item \textit{Explicit Definitions:} Universities should provide a clear definition of what constitutes acceptable AI assistance. Guidelines must specify permitted uses (e.g., brainstorming, initial drafting, language editing) and delineate prohibited practices (e.g., outsourcing entire assignments).
        \item \textit{Mandatory Disclosure:} All instances of AI-generated content must be disclosed. Just as academic work requires proper citation, the use of generative AI should be transparently noted in all submissions.
        \item \textit{Illustrative Examples:} Policy documents should include examples and case studies that illustrate acceptable and unacceptable practices, reducing ambiguity and guiding student behaviour.
    \end{enumerate}

    \end{enumerate}

Previous exhortations have focussed on having clear guidelines as to acceptable use, but the ability to actually monitor and enforce these makes them of less practical import.  Whilst clarity is important so that people are clear about what is expected, and they are the easiest action to take, they are the least actionable and effective in policy terms, and hence are presented last.

\section{Conclusion}\label{conclusion}

Generative AI, exemplified by tools like ChatGPT, is reshaping the landscape of higher education. Its transformative potential to enhance research productivity and teaching effectiveness is clear; yet, the risks it poses—especially in terms of academic integrity and equitable learning—are equally significant. Empirical evidence indicates that nearly half of students are already using LLMs in their coursework, and current detection tools are not infallible. As a result, universities must urgently adopt adaptive, transparent, and comprehensive policies to navigate this new terrain.

This article has examined the opportunities and challenges presented by generative AI, drawing on recent studies and international case examples. It has argued that the integration of AI in academia demands a multifaceted policy response—one that primarily redesigns assessments to promote original thought, enhances training for both staff and students,  implements robust enforcement mechanisms, and clearly defines acceptable use. Institutional leadership and a culture of continuous improvement will be critical to ensuring that the benefits of AI are harnessed while preserving the core values of academic rigour and integrity.

The future of higher education in the AI era will depend on our ability to balance innovation with accountability. By embracing adaptive policies and fostering an environment of transparency and ethical practice, universities can transform generative AI from a potential threat into a powerful tool for learning and discovery.

\section*{Acknowledgements}

\bibliography{references}

\begin{thebibliography}{}
\renewcommand{\doi}[1]{\url{https://doi.org/#1}}
\bibcommenthead

\bibitem [\protect \citeauthoryear {%
Bloom%
}{%
Bloom%
}{%
{\protect \APACyear {1984}}%
}]{%
bloom_2_1984}
\APACinsertmetastar {%
bloom_2_1984}%
\begin{APACrefauthors}%
Bloom, B.S.%
\end{APACrefauthors}%
\unskip\
\newblock
\APACrefYearMonthDay{1984}{}{}.
\newblock
{\BBOQ}\APACrefatitle {The 2 sigma problem: {The} search for methods of group instruction as effective as one-to-one tutoring} {The 2 sigma problem: {The} search for methods of group instruction as effective as one-to-one tutoring}.{\BBCQ}
\newblock
\APACjournalVolNumPages{Educational researcher}{13}{6}{4--16,}
\newblock
\APACrefnote{Publisher: Sage Publications Sage CA: Thousand Oaks, CA}
\newblock

\newblock

\PrintBackRefs{\CurrentBib}

\bibitem [\protect \citeauthoryear {%
Cotton%
, Cotton%
\BCBL {}\ \BBA {} Shipway%
}{%
Cotton%
\ \protect \BOthers {.}}{%
{\protect \APACyear {2024}}%
}]{%
Cotton2024}
\APACinsertmetastar {%
Cotton2024}%
\begin{APACrefauthors}%
Cotton, D.R.E.%
, Cotton, P.A.%
\BCBL {} Shipway, J.R.%
\end{APACrefauthors}%
\unskip\
\newblock
\APACrefYearMonthDay{2024}{}{}.
\newblock
{\BBOQ}\APACrefatitle {Chatting and cheating: {Ensuring} academic integrity in the era of {ChatGPT}} {Chatting and cheating: {Ensuring} academic integrity in the era of {ChatGPT}}.{\BBCQ}
\newblock
\APACjournalVolNumPages{Innovations in Education and Teaching International}{61}{2}{228--239,}
\newblock
\begin{APACrefDOI} \doi{10.1080/14703297.2023.2190148} \end{APACrefDOI}
\newblock

\newblock

\PrintBackRefs{\CurrentBib}

\bibitem [\protect \citeauthoryear {%
Dhuliawala%
\ \protect \BOthers {.}}{%
Dhuliawala%
\ \protect \BOthers {.}}{%
{\protect \APACyear {2023}}%
}]{%
dhuliawala_chain--verification_2023}
\APACinsertmetastar {%
dhuliawala_chain--verification_2023}%
\begin{APACrefauthors}%
Dhuliawala, S.%
, Komeili, M.%
, Xu, J.%
, Raileanu, R.%
, Li, X.%
, Celikyilmaz, A.%
\BCBL {} Weston, J.%
\end{APACrefauthors}%
\unskip\
\newblock
\APACrefYearMonthDay{2023}{{\APACmonth{09}}}{}.
\newblock
\APACrefbtitle {Chain-of-{Verification} {Reduces} {Hallucination} in {Large} {Language} {Models}.} {Chain-of-{Verification} {Reduces} {Hallucination} in {Large} {Language} {Models}.}
\newblock
\APACaddressPublisher{}{arXiv}.
\newblock
\begin{APACrefURL} [{2025-04-01}]{http://arxiv.org/abs/2309.11495} \end{APACrefURL}
\newblock
\APACrefnote{arXiv:2309.11495 [cs]}
\PrintBackRefs{\CurrentBib}

\bibitem [\protect \citeauthoryear {%
Freeman%
}{%
Freeman%
}{%
{\protect \APACyear {2025}}%
}]{%
HEPI2025}
\APACinsertmetastar {%
HEPI2025}%
\begin{APACrefauthors}%
Freeman, J.%
\end{APACrefauthors}%
\unskip\
\newblock
\APACrefYearMonthDay{2025}{}{}.
\newblock
\APACrefbtitle {{HEPI}/{Kortext} {AI} survey shows explosive increase in the use of generative {AI} tools by students} {{HEPI}/{Kortext} {AI} survey shows explosive increase in the use of generative {AI} tools by students}\ \APACbVolEdTR{}{\BTR{}}.
\newblock
\APACaddressInstitution{}{Higher Education Policy Institute (HEPI)}.
\newblock
\begin{APACrefURL} {https://www.hepi.ac.uk/2025/02/26/hepi-kortext-ai-survey-shows-explosive-increase-in-the-use-of-generative-ai-tools-by-students/} \end{APACrefURL}
\PrintBackRefs{\CurrentBib}

\bibitem [\protect \citeauthoryear {%
Huang%
\ \protect \BOthers {.}}{%
Huang%
\ \protect \BOthers {.}}{%
{\protect \APACyear {2025}}%
}]{%
huang_survey_2025}
\APACinsertmetastar {%
huang_survey_2025}%
\begin{APACrefauthors}%
Huang, L.%
, Yu, W.%
, Ma, W.%
, Zhong, W.%
, Feng, Z.%
, Wang, H.%
\BDBL {}Liu, T.%
\end{APACrefauthors}%
\unskip\
\newblock
\APACrefYearMonthDay{2025}{{\APACmonth{01}}}{}.
\newblock
{\BBOQ}\APACrefatitle {A {Survey} on {Hallucination} in {Large} {Language} {Models}: {Principles}, {Taxonomy}, {Challenges}, and {Open} {Questions}} {A {Survey} on {Hallucination} in {Large} {Language} {Models}: {Principles}, {Taxonomy}, {Challenges}, and {Open} {Questions}}.{\BBCQ}
\newblock
\APACjournalVolNumPages{ACM Trans. Inf. Syst.}{43}{2}{42:1--42:55,}
\newblock
\begin{APACrefDOI} \doi{10.1145/3703155} \end{APACrefDOI}
\newblock
\begin{APACrefURL} [{2025-04-01}]{https://doi.org/10.1145/3703155} \end{APACrefURL}
\newblock

\newblock

\PrintBackRefs{\CurrentBib}

\bibitem [\protect \citeauthoryear {%
Kim%
\ \protect \BOthers {.}}{%
Kim%
\ \protect \BOthers {.}}{%
{\protect \APACyear {2025}}%
}]{%
Kim2025}
\APACinsertmetastar {%
Kim2025}%
\begin{APACrefauthors}%
Kim, J.%
, Klopfer, M.%
, Grohs, J.R.%
, Eldardiry, H.%
, Weichert, J.%
, Cox, L.A.I.%
\BCBL {} Pike, D.%
\end{APACrefauthors}%
\unskip\
\newblock
\APACrefYearMonthDay{2025}{}{}.
\newblock
{\BBOQ}\APACrefatitle {Examining faculty and student perceptions of generative {AI} in university courses} {Examining faculty and student perceptions of generative {AI} in university courses}.{\BBCQ}
\newblock
\APACjournalVolNumPages{Innovative Higher Education}{}{}{online first, https://doi.org/10.1007/s10755--024--09774--w,}
\newblock
\begin{APACrefDOI} \doi{10.1007/s10755-024-09774-w} \end{APACrefDOI}
\newblock

\newblock

\PrintBackRefs{\CurrentBib}

\bibitem [\protect \citeauthoryear {%
Kinder%
\ \protect \BOthers {.}}{%
Kinder%
\ \protect \BOthers {.}}{%
{\protect \APACyear {2024}}%
}]{%
Kinder2024}
\APACinsertmetastar {%
Kinder2024}%
\begin{APACrefauthors}%
Kinder, A.%
, Briese, F.J.%
, Jacobs, M.%
, Dern, N.%
, Glodny, N.%
, Jacobs, S.%
\BCBL {} Leßmann, S.%
\end{APACrefauthors}%
\unskip\
\newblock
\APACrefYearMonthDay{2024}{}{}.
\newblock
{\BBOQ}\APACrefatitle {Effects of adaptive feedback generated by a large language model: {A} case study in teacher education} {Effects of adaptive feedback generated by a large language model: {A} case study in teacher education}.{\BBCQ}
\newblock
\APACjournalVolNumPages{Computers and Education: Artificial Intelligence}{8}{}{100349,}
\newblock
\begin{APACrefDOI} \doi{10.1016/j.caeai.2024.100349} \end{APACrefDOI}
\newblock

\newblock

\PrintBackRefs{\CurrentBib}

\bibitem [\protect \citeauthoryear {%
Korinek%
}{%
Korinek%
}{%
{\protect \APACyear {2023}}%
}]{%
Korinek2023}
\APACinsertmetastar {%
Korinek2023}%
\begin{APACrefauthors}%
Korinek, A.%
\end{APACrefauthors}%
\unskip\
\newblock
\APACrefYearMonthDay{2023}{}{}.
\newblock
{\BBOQ}\APACrefatitle {Generative {AI} for economic research: {Use} cases and implications for economists} {Generative {AI} for economic research: {Use} cases and implications for economists}.{\BBCQ}
\newblock
\APACjournalVolNumPages{Journal of Economic Literature}{61}{4}{1281--1317,}
\newblock
\begin{APACrefDOI} \doi{10.1257/jel.20231736} \end{APACrefDOI}
\newblock

\newblock

\PrintBackRefs{\CurrentBib}

\bibitem [\protect \citeauthoryear {%
Labadze%
, Grigolia%
\BCBL {}\ \BBA {} Machaidze%
}{%
Labadze%
\ \protect \BOthers {.}}{%
{\protect \APACyear {2023}}%
}]{%
labadze_role_2023}
\APACinsertmetastar {%
labadze_role_2023}%
\begin{APACrefauthors}%
Labadze, L.%
, Grigolia, M.%
\BCBL {} Machaidze, L.%
\end{APACrefauthors}%
\unskip\
\newblock
\APACrefYearMonthDay{2023}{}{}.
\newblock
{\BBOQ}\APACrefatitle {Role of {AI} {Chatbots} in {Education}: {A} {Systematic} {Literature} {Review}} {Role of {AI} {Chatbots} in {Education}: {A} {Systematic} {Literature} {Review}}.{\BBCQ}
\newblock
\APACjournalVolNumPages{International Journal of Educational Technology in Higher Education}{20}{56}{,}
\newblock
\begin{APACrefDOI} \doi{10.1186/s41239-023-00426-1} \end{APACrefDOI}
\newblock
\begin{APACrefURL} {https://doi.org/10.1186/s41239-023-00426-1} \end{APACrefURL}
\newblock

\newblock

\PrintBackRefs{\CurrentBib}

\bibitem [\protect \citeauthoryear {%
Marvin%
, Hellen%
, Jjingo%
\BCBL {}\ \BBA {} Nakatumba-Nabende%
}{%
Marvin%
\ \protect \BOthers {.}}{%
{\protect \APACyear {2024}}%
}]{%
marvin_prompt_2024}
\APACinsertmetastar {%
marvin_prompt_2024}%
\begin{APACrefauthors}%
Marvin, G.%
, Hellen, N.%
, Jjingo, D.%
\BCBL {} Nakatumba-Nabende, J.%
\end{APACrefauthors}%
\unskip\
\newblock
\APACrefYearMonthDay{2024}{}{}.
\newblock
{\BBOQ}\APACrefatitle {Prompt {Engineering} in {Large} {Language} {Models}} {Prompt {Engineering} in {Large} {Language} {Models}}.{\BBCQ}
\newblock
 I.J.~Jacob, S.~Piramuthu\BCBL {}\ \BBA {} P.~Falkowski-Gilski\ (\BEDS), \APACrefbtitle {Data {Intelligence} and {Cognitive} {Informatics}} {Data {Intelligence} and {Cognitive} {Informatics}}\ (\BPGS\ 387--402).
\newblock
\APACaddressPublisher{Singapore}{Springer Nature}.
\PrintBackRefs{\CurrentBib}

\bibitem [\protect \citeauthoryear {%
Mehrabi%
, Morstatter%
, Saxena%
, Lerman%
\BCBL {}\ \BBA {} Galstyan%
}{%
Mehrabi%
\ \protect \BOthers {.}}{%
{\protect \APACyear {2021}}%
}]{%
mehrabi_survey_2021}
\APACinsertmetastar {%
mehrabi_survey_2021}%
\begin{APACrefauthors}%
Mehrabi, N.%
, Morstatter, F.%
, Saxena, N.%
, Lerman, K.%
\BCBL {} Galstyan, A.%
\end{APACrefauthors}%
\unskip\
\newblock
\APACrefYearMonthDay{2021}{}{}.
\newblock
{\BBOQ}\APACrefatitle {A {Survey} on {Bias} and {Fairness} in {Machine} {Learning}} {A {Survey} on {Bias} and {Fairness} in {Machine} {Learning}}.{\BBCQ}
\newblock
\APACjournalVolNumPages{ACM Computing Surveys}{54}{6}{1--35,}
\newblock
\begin{APACrefDOI} \doi{10.1145/3457607} \end{APACrefDOI}
\newblock

\newblock

\PrintBackRefs{\CurrentBib}

\bibitem [\protect \citeauthoryear {%
{Monash University}%
}{%
{Monash University}%
}{%
{\protect \APACyear {{\protect \bibnodate {}}}}%
}]{%
monash_university_ai_nodate}
\APACinsertmetastar {%
monash_university_ai_nodate}%
\begin{APACrefauthors}%
{Monash University}%
\end{APACrefauthors}%
\unskip\
\newblock
\APACrefYearMonthDay{{\protect \bibnodate {}}}{}{}.
\newblock
\APACrefbtitle {{AI} and assessment.} {{AI} and assessment.}
\newblock
\begin{APACrefURL} [{2025-04-25}]{https://www.monash.edu/learning-teaching/teachhq/Teaching-practices/artificial-intelligence/ai-and-assessment} \end{APACrefURL}
\PrintBackRefs{\CurrentBib}

\bibitem [\protect \citeauthoryear {%
of Universities%
}{%
of Universities%
}{%
{\protect \APACyear {2023}}%
}]{%
noauthor_russell_2023}
\APACinsertmetastar {%
noauthor_russell_2023}%
\begin{APACrefauthors}%
of Universities, R.G.%
\end{APACrefauthors}%
\unskip\
\newblock
\APACrefYearMonthDay{2023}{}{}.
\newblock
\APACrefbtitle {Russell {Group} principles on the use of generative {AI} tools in education.} {Russell {Group} principles on the use of generative {AI} tools in education.}
\newblock
\begin{APACrefURL} [{2025-04-03}]{https://www.russellgroup.ac.uk/sites/default/files/2025-01/Russell\%20Group\%20principles\%20on\%20generative\%20AI\%20in\%20education.pdf} \end{APACrefURL}
\PrintBackRefs{\CurrentBib}

\bibitem [\protect \citeauthoryear {%
Page%
\ \protect \BOthers {.}}{%
Page%
\ \protect \BOthers {.}}{%
{\protect \APACyear {2021}}%
}]{%
page_prisma_2021}
\APACinsertmetastar {%
page_prisma_2021}%
\begin{APACrefauthors}%
Page, M.J.%
, McKenzie, J.E.%
, Bossuyt, P.M.%
, Boutron, I.%
, Hoffmann, T.C.%
, Mulrow, C.D.%
\BDBL {}Moher, D.%
\end{APACrefauthors}%
\unskip\
\newblock
\APACrefYearMonthDay{2021}{{\APACmonth{03}}}{}.
\newblock
{\BBOQ}\APACrefatitle {The {PRISMA} 2020 statement: an updated guideline for reporting systematic reviews} {The {PRISMA} 2020 statement: an updated guideline for reporting systematic reviews}.{\BBCQ}
\newblock
\APACjournalVolNumPages{BMJ}{372}{}{n71,}
\newblock
\begin{APACrefDOI} \doi{10.1136/bmj.n71} \end{APACrefDOI}
\newblock
\begin{APACrefURL} [{2025-04-02}]{https://www.bmj.com/content/372/bmj.n71} \end{APACrefURL}
\newblock
\APACrefnote{Publisher: British Medical Journal Publishing Group Section: Research Methods \&amp; Reporting}
\newblock

\newblock

\PrintBackRefs{\CurrentBib}

\bibitem [\protect \citeauthoryear {%
Paustian%
\ \BBA {} Slinger%
}{%
Paustian%
\ \BBA {} Slinger%
}{%
{\protect \APACyear {2024}}%
}]{%
Paustian2024}
\APACinsertmetastar {%
Paustian2024}%
\begin{APACrefauthors}%
Paustian, T.%
\BCBT {}\ \BBA {} Slinger, B.%
\end{APACrefauthors}%
\unskip\
\newblock
\APACrefYearMonthDay{2024}{}{}.
\newblock
{\BBOQ}\APACrefatitle {Students are using large language models and {AI} detectors can often detect their use} {Students are using large language models and {AI} detectors can often detect their use}.{\BBCQ}
\newblock
\APACjournalVolNumPages{Frontiers in Education}{9}{}{1374889,}
\newblock
\begin{APACrefDOI} \doi{10.3389/feduc.2024.1374889} \end{APACrefDOI}
\newblock

\newblock

\PrintBackRefs{\CurrentBib}

\bibitem [\protect \citeauthoryear {%
Phoenix%
\ \BBA {} Taylor%
}{%
Phoenix%
\ \BBA {} Taylor%
}{%
{\protect \APACyear {2024}}%
}]{%
phoenix2024prompt}
\APACinsertmetastar {%
phoenix2024prompt}%
\begin{APACrefauthors}%
Phoenix, J.%
\BCBT {}\ \BBA {} Taylor, M.%
\end{APACrefauthors}%
\unskip\
\newblock
\APACrefYear{2024}.
\newblock
\APACrefbtitle {Prompt engineering for generative {AI}} {Prompt engineering for generative {AI}}.
\newblock
\APACaddressPublisher{}{" O'Reilly Media, Inc."}.
\PrintBackRefs{\CurrentBib}

\bibitem [\protect \citeauthoryear {%
{Quality}%
\ \BBA {} {Agency}%
}{%
{Quality}%
\ \BBA {} {Agency}%
}{%
{\protect \APACyear {2025}}%
}]{%
noauthor_artificial_nodate}
\APACinsertmetastar {%
noauthor_artificial_nodate}%
\begin{APACrefauthors}%
{Quality}, T.E.%
\BCBT {}\ \BBA {} {Agency}, s.%
\end{APACrefauthors}%
\unskip\
\newblock
\APACrefYearMonthDay{2025}{}{}.
\newblock
\APACrefbtitle {Artificial intelligence {\textbar} {Tertiary} {Education} {Quality} and {Standards} {Agency}.} {Artificial intelligence {\textbar} {Tertiary} {Education} {Quality} and {Standards} {Agency}.}
\newblock
\begin{APACrefURL} [{2025-04-25}]{https://www.teqsa.gov.au/guides-resources/higher-education-good-practice-hub/artificial-intelligence} \end{APACrefURL}
\PrintBackRefs{\CurrentBib}

\bibitem [\protect \citeauthoryear {%
Razafinirina%
, Dimbisoa%
\BCBL {}\ \BBA {} Mahatody%
}{%
Razafinirina%
\ \protect \BOthers {.}}{%
{\protect \APACyear {2024}}%
}]{%
razafinirina_pedagogical_2024}
\APACinsertmetastar {%
razafinirina_pedagogical_2024}%
\begin{APACrefauthors}%
Razafinirina, M.A.%
, Dimbisoa, W.G.%
\BCBL {} Mahatody, T.%
\end{APACrefauthors}%
\unskip\
\newblock
\APACrefYearMonthDay{2024}{}{}.
\newblock
{\BBOQ}\APACrefatitle {Pedagogical {Alignment} of {Large} {Language} {Models} ({LLM}) for {Personalized} {Learning}: {A} {Survey}, {Trends} and {Challenges}} {Pedagogical {Alignment} of {Large} {Language} {Models} ({LLM}) for {Personalized} {Learning}: {A} {Survey}, {Trends} and {Challenges}}.{\BBCQ}
\newblock
\APACjournalVolNumPages{Journal of Intelligent Learning Systems and Applications}{16}{4}{--,}
\newblock
\begin{APACrefDOI} \doi{10.4236/jilsa.2024.164009} \end{APACrefDOI}
\newblock
\begin{APACrefURL} {https://doi.org/10.4236/jilsa.2024.164009} \end{APACrefURL}
\newblock

\newblock

\PrintBackRefs{\CurrentBib}

\bibitem [\protect \citeauthoryear {%
Seckel%
, Stephens%
\BCBL {}\ \BBA {} Rodriguez%
}{%
Seckel%
\ \protect \BOthers {.}}{%
{\protect \APACyear {2024}}%
}]{%
Seckel2024}
\APACinsertmetastar {%
Seckel2024}%
\begin{APACrefauthors}%
Seckel, E.%
, Stephens, B.Y.%
\BCBL {} Rodriguez, F.%
\end{APACrefauthors}%
\unskip\
\newblock
\APACrefYearMonthDay{2024}{}{}.
\newblock
{\BBOQ}\APACrefatitle {Ten simple rules to leverage large language models for getting grants} {Ten simple rules to leverage large language models for getting grants}.{\BBCQ}
\newblock
\APACjournalVolNumPages{PLoS Computational Biology}{20}{3}{e1011863,}
\newblock
\begin{APACrefDOI} \doi{10.1371/journal.pcbi.1011863} \end{APACrefDOI}
\newblock

\newblock

\PrintBackRefs{\CurrentBib}

\bibitem [\protect \citeauthoryear {%
Shah%
\ \protect \BOthers {.}}{%
Shah%
\ \protect \BOthers {.}}{%
{\protect \APACyear {2024}}%
}]{%
shah_students_2024}
\APACinsertmetastar {%
shah_students_2024}%
\begin{APACrefauthors}%
Shah, M.%
, Pankiewicz, M.%
, Baker, R.S.%
, Chi, J.%
, Xin, Y.%
, Shah, H.%
\BCBL {} Fonseca, D.%
\end{APACrefauthors}%
\unskip\
\newblock
\APACrefYearMonthDay{2024}{{\APACmonth{11}}}{}.
\newblock
{\BBOQ}\APACrefatitle {Students’ {Use} of an {LLM}-{Powered} {Virtual} {Teaching} {Assistant} for {Recommending} {Educational} {Applications} of {Games}} {Students’ {Use} of an {LLM}-{Powered} {Virtual} {Teaching} {Assistant} for {Recommending} {Educational} {Applications} of {Games}}.{\BBCQ}
\newblock
 \APACrefbtitle {Serious {Games}: 10th {Joint} {International} {Conference}, {JCSG} 2024, {New} {York} {City}, {NY}, {USA}, {November} 7–8, 2024, {Proceedings}} {Serious {Games}: 10th {Joint} {International} {Conference}, {JCSG} 2024, {New} {York} {City}, {NY}, {USA}, {November} 7–8, 2024, {Proceedings}}\ (\BPGS\ 19--24).
\newblock
\APACaddressPublisher{Berlin, Heidelberg}{Springer-Verlag}.
\newblock
\begin{APACrefURL} [{2025-04-03}]{https://doi.org/10.1007/978-3-031-74138-8\_2} \end{APACrefURL}
\PrintBackRefs{\CurrentBib}

\bibitem [\protect \citeauthoryear {%
Tang%
, Duan%
\BCBL {}\ \BBA {} Cai%
}{%
Tang%
\ \protect \BOthers {.}}{%
{\protect \APACyear {2024}}%
}]{%
Tang2024}
\APACinsertmetastar {%
Tang2024}%
\begin{APACrefauthors}%
Tang, X.%
, Duan, X.%
\BCBL {} Cai, Z.G.%
\end{APACrefauthors}%
\unskip\
\newblock
\APACrefYearMonthDay{2024}{}{}.
\newblock
\APACrefbtitle {Are {LLMs} good literature review writers? {Evaluating} the literature review writing ability of large language models.} {Are {LLMs} good literature review writers? {Evaluating} the literature review writing ability of large language models.}
\newblock
\APACrefnote{tex.howpublished: arXiv preprint arXiv:2412.13612}
\PrintBackRefs{\CurrentBib}

\end{thebibliography}

\end{document}